\documentclass[a4paper,12pt]{article}
\usepackage[utf8]{inputenc}
\usepackage{jheppub}
\usepackage{tikz}
\usepackage[miktex]{gnuplottex}
\usetikzlibrary{shapes,arrows,positioning,automata,backgrounds,calc,er,patterns}
\usepackage{tikz-feynman}
\tikzfeynmanset{compat=1.0.0}
\usepackage{soul}
\usepackage{slashed} 
\usepackage{enumerate}
\usepackage{epsfig} 
\usepackage{float} 
\usepackage{subcaption} 
\usepackage{amsmath} 
\usepackage{wasysym} 
\usepackage{mathrsfs} 
\usepackage{amsfonts} 
\usepackage{amsbsy} 
\usepackage{amscd}    
\usepackage{multirow} 
\usepackage{tikz}
\usepackage{slashed}
\usepackage[miktex]{gnuplottex}
\usetikzlibrary{arrows,positioning,shapes.geometric} 
\usepackage{color}
\usepackage{verbatim}
\usepackage{xcolor}
\usepackage{aligned-overset}

\makeatletter

\def\eeq{\end{equation}}
\def\beq{\begin{equation}}

\newcommand{\Rmnum}[1]{\expandafter\@slowromancap\romannumeral #1@}

\newcommand{\bea} {\begin{eqnarray}}
\newcommand{\eea} {\end{eqnarray}}

\newcommand{\gsim}{\raisebox{-0.13cm}{~\shortstack{$>$ \\[-0.07cm]
      $\sim$}}~}
\newcommand{\lsim}{\raisebox{-0.13cm}{~\shortstack{$<$ \\[-0.07cm]
      $\sim$}}~}
\makeatother

\title{Uncovering TeV-scale dark matter with Z-fusion at a muon collider}

\author[a]{Anindya Datta,}
\author[b]{Jayita Lahiri,}
\author[b,c]{Biswarup Mukhopadhyaya}

\affiliation[a]{Department of Physics, University of Calcutta, 92 Acharya Prafulla Chandra Road, Kolkata 700009, India}

\affiliation[b]{Department of Physical Sciences, Indian Institute of Science Education and Research Kolkata, Mohanpur - 741246, India}

\affiliation[c]{School of Physical Sciences, Indian Association for the Cultivation of Science, 2A \protect\& 2B Raja S. C. Mallick Road, Jadavpur, Kolkata - 700032, India.}

\emailAdd{adphys@caluniv.ac.in}

\emailAdd{jayitalahiri@iiserkol.ac.in}

\emailAdd{biswarup.mukho@gmail.com}

\abstract{We demonstrate the advantage of the $Z$-boson fusion channel at a high-energy, high-luminosity muon collider in eliciting
signals of a TeV-scale scalar dark matter (DM) particle.
The utility of such a set-up lies in (a)the attainability
of high centre-of-mass energy ($\approx 10 ~\rm TeV$), (b)the
likely availability of large integrated luminosity 
($\gsim 10~ ab^{-1}$), and (c)the facility for efficient detection of very forward tagged muons. Our analysis is based on the process $\mu^+ \mu^- \rightarrow \mu^+ \mu^-$ + missing energy, driven by $Z$-fusion. We present the results
in the context of (i) a scenario characterised by dimension-6 effective couplings, and (ii) an inert doublet model. After filtering each scenario through its associated constraints, we find that it is possible to achieve a signal
significance of about 6.7$\sigma$ in case (i) and about 5.5$\sigma$ in case (ii), for DM particle masses 
$\approx$ 1 TeV and an integrated luminosity of  $10 ~ ab^{-1}$.}

\begin{document}

\maketitle

\section{Introduction}

An important goal of accelerator-based particle physics is to advance 
 (a) the energy frontier, and (b) the luminosity frontier. While aspect (a) enables us
 to look for  progressively higher-mass particles, we can gain ascendancy
 in terms of  statistical significance, thus building up to prospects of `cleaner' new signals, via
 aspect (b). Therefore, a high-energy, high-luminosity lepton collider can in principle
 serve as an asset, when it comes to unravelling high-mass modes that are harbingers  of physics
 beyond the standard model (BSM), especially in the electroweak sector. Muons 
 fit into such a role better than electrons, since they associate less electromagnetic radiation loss. Also,
 their lifetime gets dilated at trans-TeV energies.
 It is in this spirit that  the present study examines the potential of a $\sqrt {s}\gsim 10$ TeV muon collider~\cite{InternationalMuonCollider:2025sys,InternationalMuonCollider:2024jyv,Accettura:2023ked,Black:2022cth,Han:2022ubw} for
 exploring (weakly interacting) trans-TeV dark matter candidates, utilising a channel that yields substantial
event rates by virtue of integrated luminosities $\gsim 10~ab^{-1}$. 

Vector boson fusion (VBF) turns out to be the most useful way of BSM search at a muon collider. Recent works have looked for dark matter (DM) candidates via this process \cite {Han:2020uak,Dutta:2025nmy}, including a multitude
of sub-processes and final states. We, on the other hand, concentrate here on a particular VBF channel,
namely, the Z-fusion process, which makes the final-state muons visible, in close resemblance to
VBF at hadron colliders~\footnote{At the LHC, the production rate of a TeV scale DM in the VBF channel turns out to be rather small.}. Though this {\it prima facie} restricts us to smaller event rates than is available in the W-fusion channel, it has the following advantages:\\

\begin{itemize}
\item For a DM candidate $\chi$, the lowest-order production mode is\\
$\mu^+\mu^- \rightarrow \mu^+\mu^- \chi \chi$, \\
leading to two energetic, high-rapidity muons in opposite directions, along with missing energy and momentum,
with the latter pointing in a direction within the  rapidity gap. The Z-fusion channel thus makes it possible to use the forward muon tags.

\item The standard model (SM) backgrounds are rather easily suppressed with the help of cuts analogous to those applied at hadron colliders, with the further advantage that the signal is free from backgrounds such as those coming from pomeron exchange. 

\item The  paucity of Z-fusion events as compared to W-fusion is compensated by the large integrated luminosity of the order of
$10~ab^{-1}$, which can go further up, if one can afford higher power consumption than a few $ab^{-1} TWh^{-1}$
for centre-of-mass energies $\gsim 10$ TeV. This happens due to the progress of the luminosity frontier 
envisioned for a muon collider.

\item Z-fusion processes lead to a large number of `very forward muons' with $|\eta| \rightarrow 7$. A dedicated
forward muon detector like MUSIC \cite{Andreetto:2025mrd,Andreetto:2026gvm} can thus enhance event rates.  One can further note that the
higher is the centre-of-mass energy of collision (which raises the mass reach for $\chi$), the event rate increases
drastically in the rapidity range $|\eta| \simeq 5-7$ for the final-state muons. Therefore, the refinement of 
high-energy muon identification in very forward directions is  crucial in utilising the high luminosity of the machine. The utility of a forward muon detector has been explored in earlier works~\cite{Ruhdorfer:2023uea,Ruhdorfer:2024dgz} . 

\item The VBF-cuts not only establish the genuineness of the signal events but also keep away beam-induced
backgrounds (BIB) which may pose problems for W-fusion events.

\end{itemize}

In this work, we demonstrate the usefulness of Z-fusion, using as illustration 
 a neutral spin-zero DM candidate $\chi$ which
is odd under a $Z_2$-symmetry, at a muon collider. We have started by considering some effective interactions driving the Z-fusion-induced production process
$\mu^+\mu^- \rightarrow \mu^+\mu^- \chi \chi$, where $\chi$ may either be an $SU(2)$ singlet or belong to
a non-trivial representation of the gauge group. Subsequently, we have also illustrated our search strategy
in the context of an inert Higgs doublet model (IDM), a rather extensively studied scenario that is otherwise
subject to strong  constraints, but less so when the DM candidate lies in the trans-TeV mass range. While the
former scenario leads to a relatively model-independent perspective, the latter brings out how some features of
the mutually related model parameters influence our predictions. However, we emphasize that our central idea
as well as the proposed search strategy is independent of this illustrative model.

The energy of each colliding muon beam is assumed to be fixed in our analysis. Although it is realistic to allow
beam energy spread of a few per cent, such spread is unlikely to have any qualitative effect on the predictions,
particularly when no mass reconstruction is attempted.

The study based on effective interactions in presented in Section 2. In Section 3, we take up an analysis
based on the IDM scenario. We summarise and conclude in Section 4.

\section{Model-independent analysis in terms of effective operators}

As has been already stated, we aim to simulate DM-pair production via Z-fusion in $\mu^+ \mu^-$ collision 
at centre-of-mass energies $\simeq 10$ TeV. It admits of two approaches: (a) when the DM candidate $\chi$ is an SU(2) singlet and (b) when it is a non-singlet. 
For case (b), one requires some theoretical model as template, which sets forth the role of the particle spectrum in the production mechanism. Then the theoretical and phenomenological constraints on the spectrum unavoidably 
play a role in the analysis of events.   This is studied in the subsequent section, after taking up case (a) first, where  
one can use effective interaction terms involving $\chi$ in a purely phenomenological manner, and  the investigation is relatively less constrained.

\subsection{Effective interactions : The process and its constraints}

Since our desired process(es) are driven by VBF,
we postulate a set of illustrative dimension-6  effective interactions between DM-pairs and gauge boson pairs, of the form\\

\begin{equation}
{\cal L}_{int} = \frac{c_1}{\Lambda^2} B_{\mu\nu} B^{\mu\nu} \chi^2 + \frac{c_2}{\Lambda^2} W_{\mu\nu}^a W^{\mu\nu a} \chi^2
\end{equation}

\noindent
where $\chi$ is an electroweak gauge-singlet scalar odd under a postulated $Z_2$-symmetry. $\Lambda$ here is the scale the effective interaction. For the generation of the model files we use \texttt{FeynRules}~\cite{Christensen:2008py}. Keeping the energy reach of a muon collider in mind, we use $\Lambda = 12$ TeV. 

\medskip
\begin{figure}[!hptb]
    \centering
        \includegraphics[width=8.0cm,height=7.0cm]{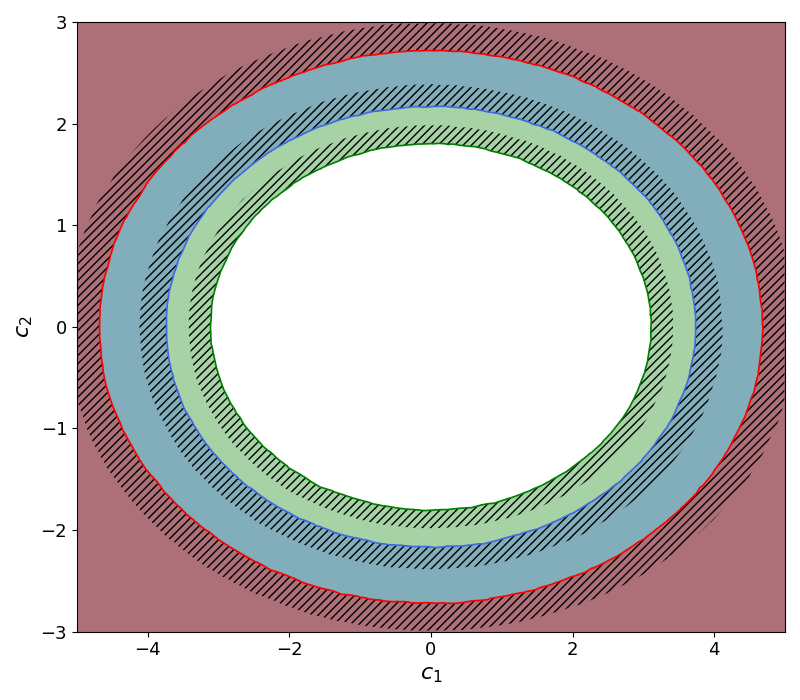}
    
    \caption{Relic under-abundant regions: purple region underabundant for $m_{\chi}=800$ GeV, purple, blue regions underabundant for $m_{\chi}=1$ TeV, purple, blue, green region underabundant for $m_{\chi}=1.2$ TeV. The hatched regions in each cases denote the regions with the observed relic abundance $\Omega h^2 = 0.12$ within 2$\sigma$. }
    \label{relic}
\end{figure}

In this scenario, the elastic scattering of DM off the nucleon takes place only via one-loop diagrams, which receives additional suppression at the coupling strength owing to very high value of $\Lambda$. Therefore, such a scenario can evade stringent constraints from direct detection experiments.

The observed relic density puts significant restrictions on the parameter space. We calculate DM observables including relic density for our parameter space using~\texttt{micrOmegas-v7.1.10}~\cite{Belanger:2026asz}. Fig.~\ref{relic} shows, for
each DM mass, the region spanned by the Wilson coefficients $c_1$ and $c_2$, which produces relic density at or below the observed value, i.e. $\Omega h^2 \lsim \Omega_{obs}h^2$, where $\Omega_{obs} h^2 = 0.12 \pm 0.0010$~\cite{Aghanim:2018eyx}.  The purple annular region corresponds to under-abundance for $m_{\chi}=800$ GeV. The purple and blue regions indicate under-abundant relic regions for $m_{\chi} = 1$ TeV. Finally, the purple, blue, and green regions correspond to relic density below the observed relic, for $m_{\chi} = 1.2$ TeV. The hatched annular regions  with the three different colours saturate the observed relic density within $2\sigma$ for the corresponding DM masses. One should note that with increasing mass of DM relic density decreases, for a given $c_1, c_2$.

We checked that the parameter space is allowed by the upper limits on DM pair annihilation cross-section from Fermi-LAT~\cite{Fermi-LAT:2011vow,Fermi-LAT:2016uux} experiment.
It is worth mentioning here that the annihilation of DM pairs into photons and charged particle pairs may lead to additional constraints from re-ionization~\cite{Slatyer:2015jla}. However, such constraints may not be binding if the $\chi$-pair annihilates into other channels as well, and also if $\chi$ is not the  overwhelmingly dominant contributor to the observed relic density. The benchmark parameters used in Fig.~\ref{relic}, including those that saturate the
observed relic density, have been found to be consistent with all such
`indirect' constraints.

We also add here that a dimension-4 Higgs portal interaction term between DM $\chi$ and Higgs pair ($\Phi^\dagger\Phi\chi^2$) is possible. However coefficients of such a term is strongly constrained by direct detection experiments since it introduces t-channel DM-nucleon scattering process at the tree-level. Furthermore, the aforementioned re-ionization constraints also restrict such terms even further. In this work, we have assumed the coefficient of the Higgs-portal term to be zero.

\subsection{Kinematical distributions}

 \begin{figure}[!hptb]
    \centering
        \includegraphics[width=6.0cm,height=5.0cm]{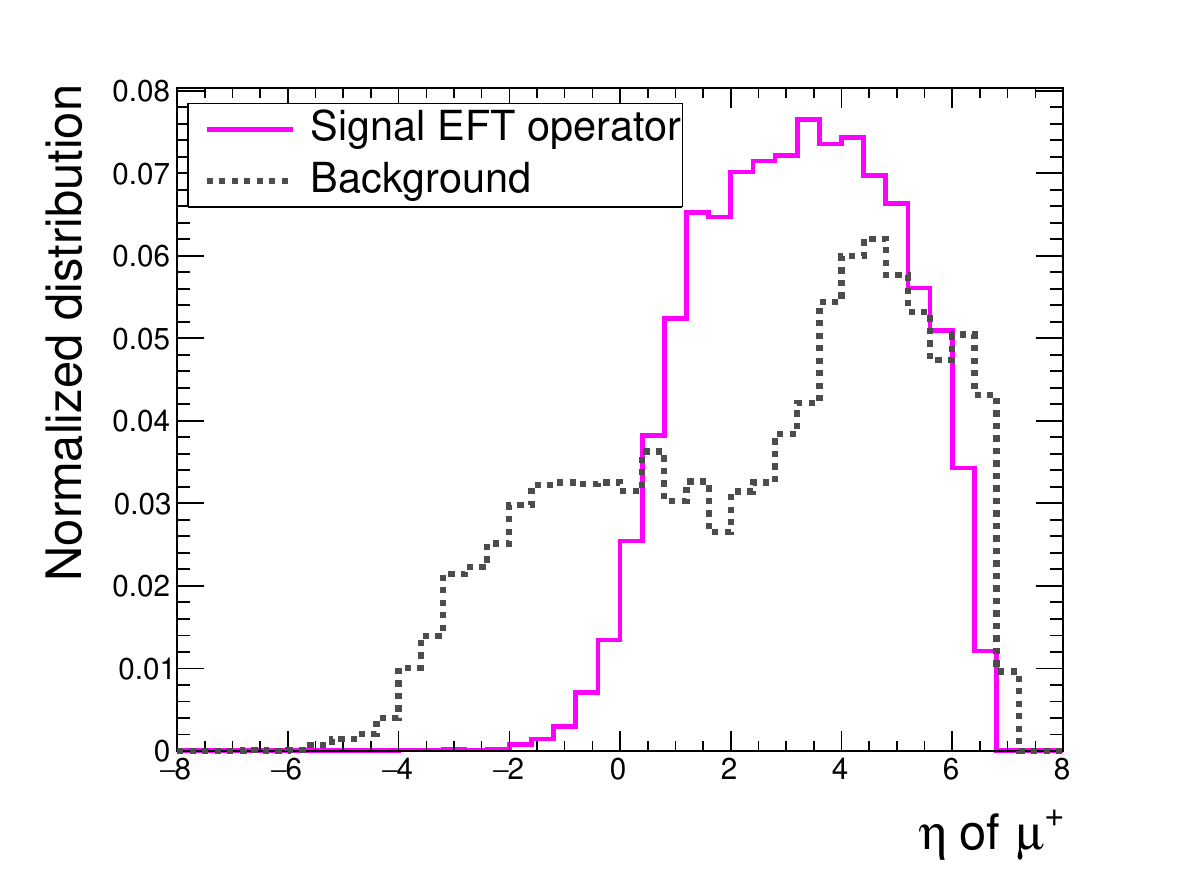}
    \includegraphics[width=6.0cm,height=5.0cm]{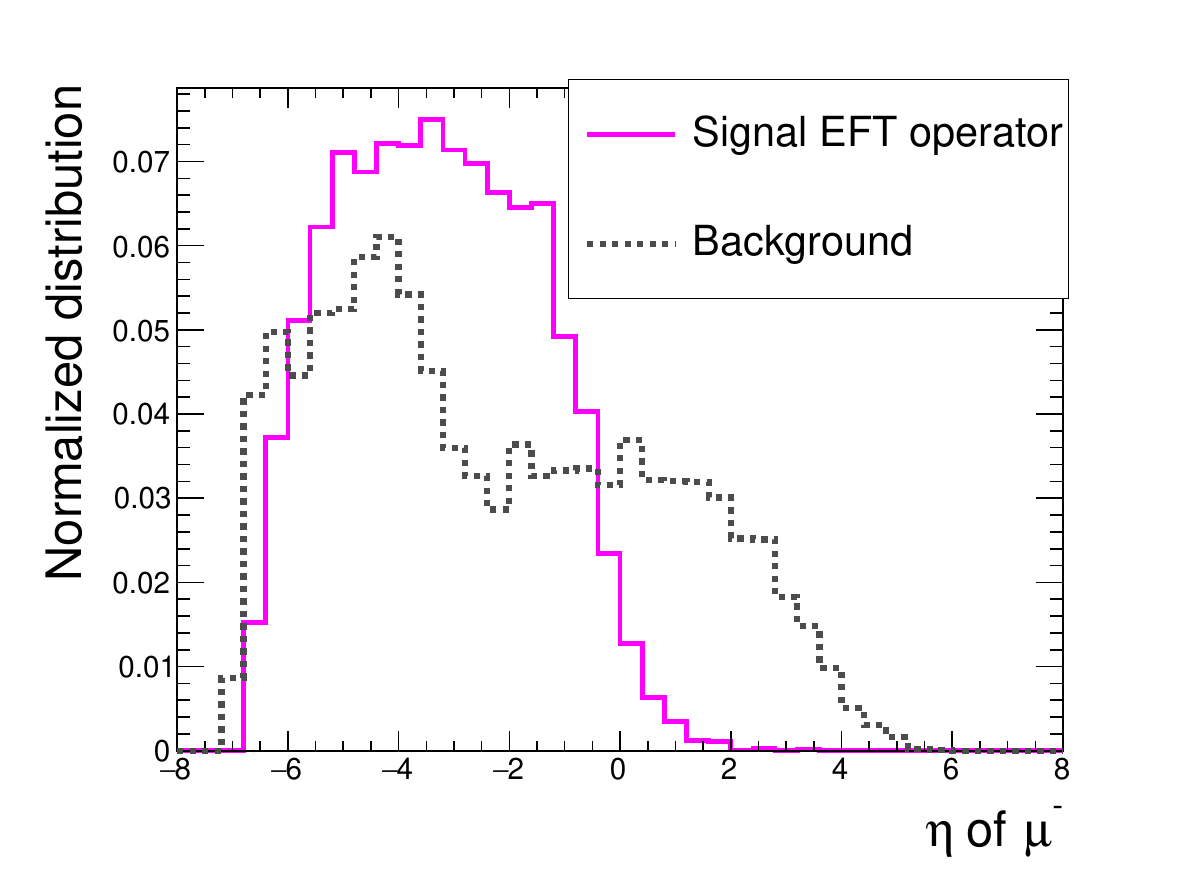}\\
    \includegraphics[width=6.0cm,height=5.0cm]{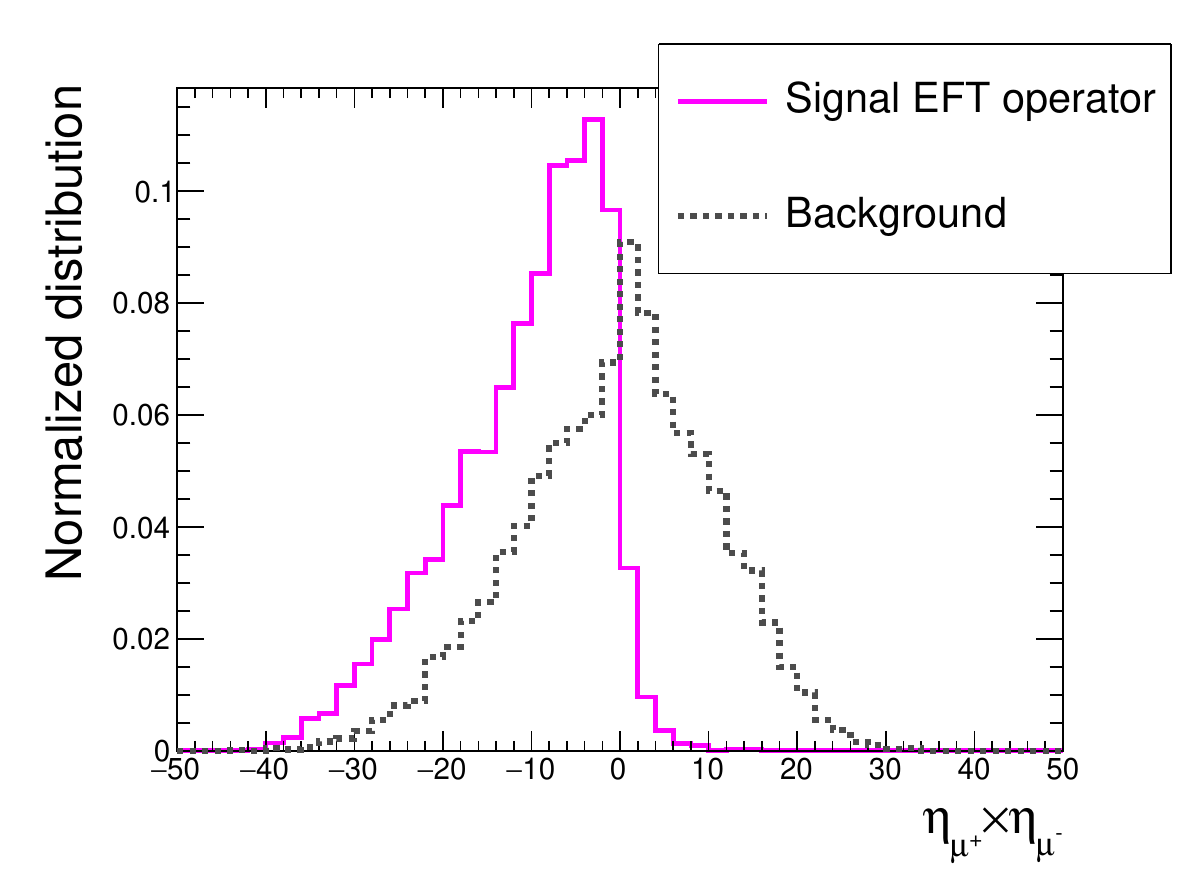}
    \includegraphics[width=6.0cm,height=5.0cm]{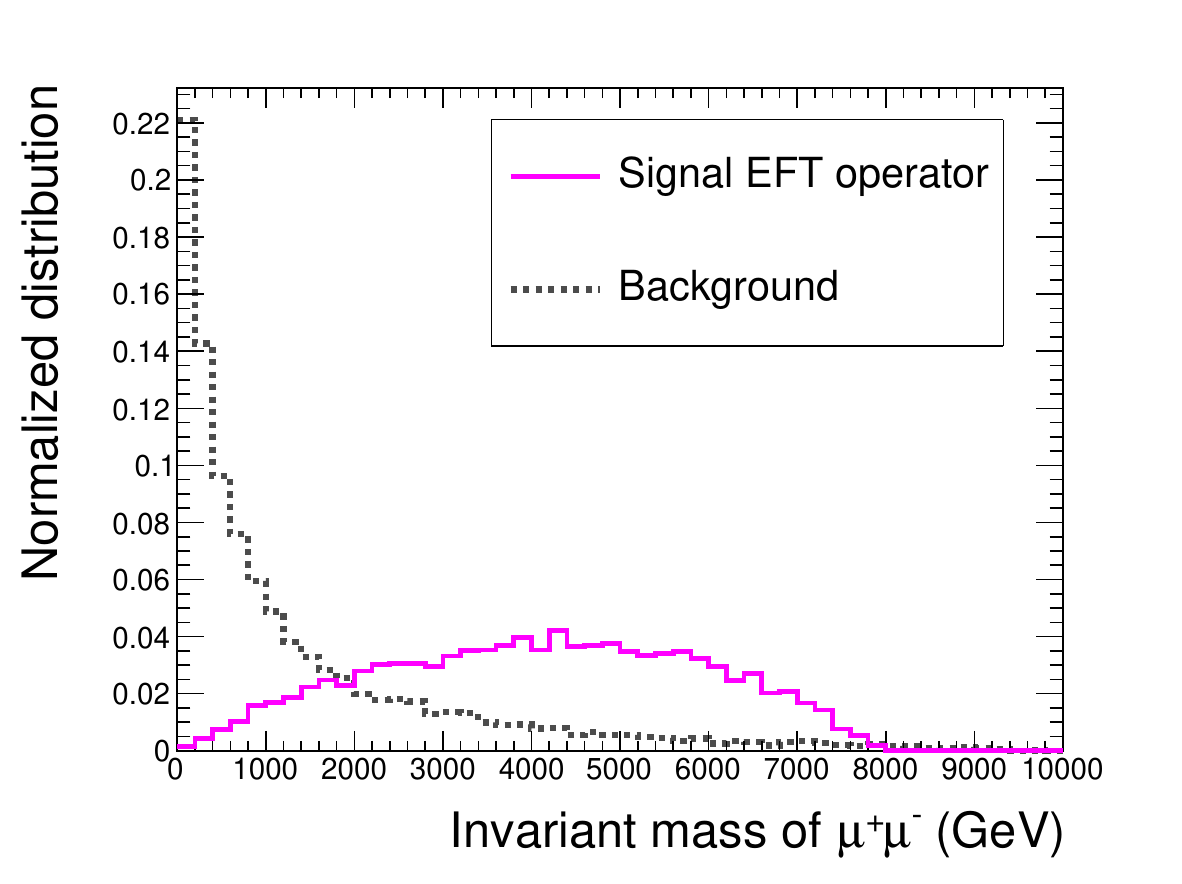}\\
    \includegraphics[width=6.0cm,height=5.0cm]{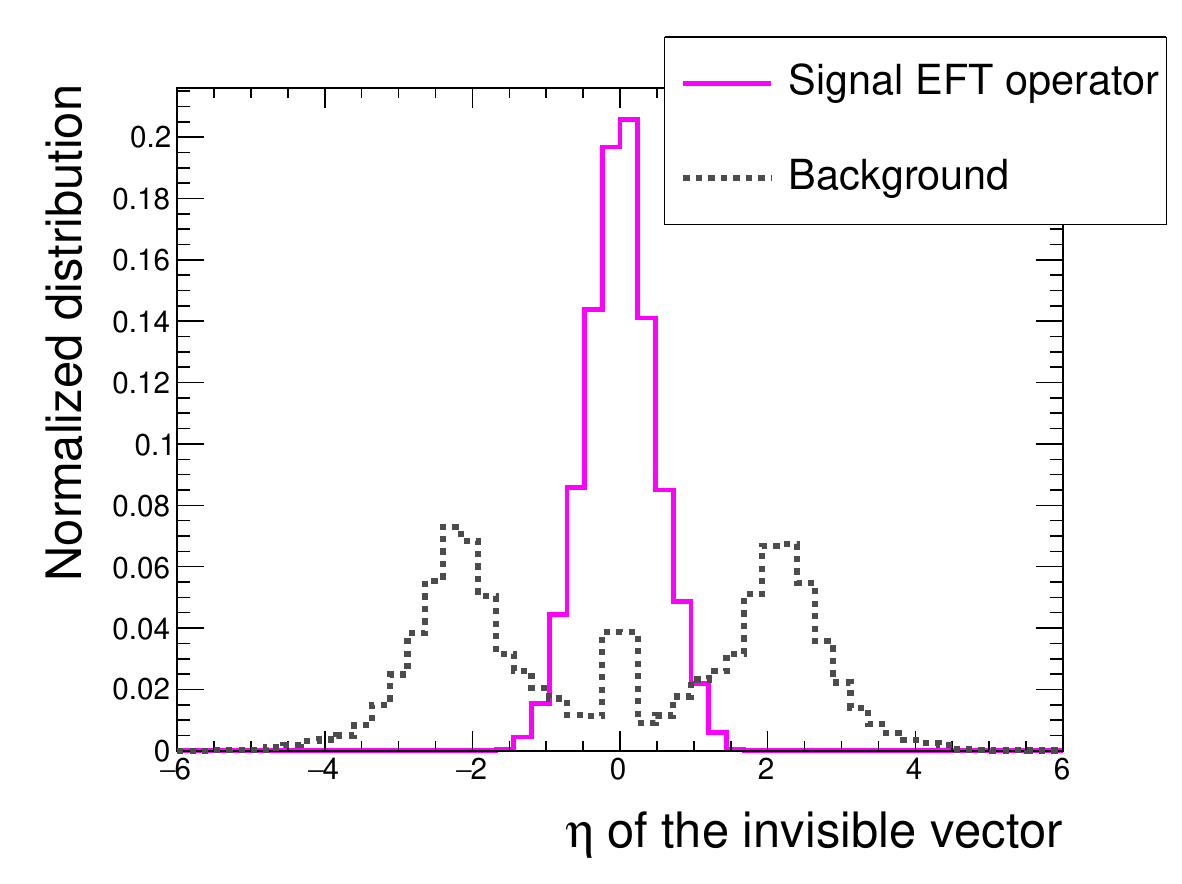}  
    \caption{Top panel: pseudo-rapidity ($\eta$) distribution of $\mu^+$ (left) and $\mu^-$ (right). Middle panel: product of pseudo-rapidity distribution of $\mu^+$ and $\mu^-$ (left), invariant mass distribution of $\mu^+$ and $\mu^-$(right). Bottom panel: rapidity distribution of the invisible vector. We have considered $\sqrt{s} = 10$ TeV. }
    \label{dist}
\end{figure}

At the muon collider, we generate our signal : two forward muons and missing energy, i.e. $\mu^+\mu^- > \chi \chi \mu^+\mu^-$. The major SM background contributing to the same final state comes from $\mu^+\mu^- > \nu\nu \mu^+\mu^-$. The signal process is mediated by vector boson fusion i.e. $ZZ, \gamma\gamma$ and $\gamma Z$-fusion. One should note that in the limit $c_1 = c_2$, the $\gamma Z$-fusion vanishes from gauge-invariance. 
We show next the kinematical distributions that help us separate signal from background. 
In Fig.~\ref{dist}, we first show the pseudo-rapidity distribution of the muons. Since our signal topology is that of vector-boson fusion, the muons are expected to be in the forward region. However, one thing to note here is that, for our dimension-6 operator in consideration, the coupling strength of the $VV^{(')}\chi\chi$ vertex is momentum-dependent. Therefore, events with higher momentum transfer through the t-channel gauge-bosons contribute more to the total amplitude. Such high momentum transfer will lead to more central muons compared to a typical VBF signal, where momentum-dependent couplings do not appear. However, the signal distribution (pink) exhibits on the whole more forward behavior compared to the background (black).

Next, we show the invariant mass between the muon pair. The signal clearly shows a higher invariant mass distribution compared to the background. Furthermore, since the two muons are expected to be in the opposite hemisphere owing to the VBF topology, the product of their $\eta$-values are expected to be negative. We show the distribution of $\eta_{\mu^-}\times \eta_{\mu^+}$, which is dominantly negative for signal. 

Finally we construct an observable from the resultant of the invisible vector. We call it y of the invisible vector ($\eta_{invisible})$ defined as follows:

\begin{equation}
\eta_{invisible} = \frac{1}{2}{\text{log}}\left[\frac{E+p_z}{E-p_z}\right]
\end{equation}

\noindent
 where $E$, $p_z$ are the energy and $z$- component of the invisible four-momentum respectively. Muon collider being a lepton machine with fixed initial centre-of-mass energy, both of these quantities can be calculated from the observed energy and momenta of the two muons.
 We show the distribution of this observable for both signal and background and we can see signal is largely concentrated around $|\eta_{invisible}|\approx 0$, whereas the background distribution is much more wide-spread. Physically this means the DM pair is always between the two forward muons and concentrated in the central region for signal. This observable proves to be the most efficient discriminator between signal and background.

\begin{table}[h]
    \centering
    \begin{tabular}{c|c|c}
       Cuts  & No of signal events &  No of Background events \\
        \hline
        Selection cuts  &  1025  &  $1.44\times 10^7$ \\
         \hline
         2 isolated muons  & 1023 & $1.44\times 10^7$ \\
        \hline
        $1 < |\eta_{\mu}|_{1,2} < 7$ & 747 & $8.9 \times 10^6$\\
        \hline
        $\eta_{\mu^+}\times\eta_{\mu^-} < 0$ & 741 & $4.4 \times 10^6$ \\
        \hline
        Invariant mass of $\mu^+\mu^- > 1$ TeV & 729 & $4.1 \times 10^6$\\
        \hline
        $|\eta_{invisible}| < 0.5$ & 511 & 5760\\
        \hline
        
   \end{tabular}
   \caption{Cut flow for signal $m_\chi=1$ TeV, $c_1=2, c_2=2, \Lambda = 12$ TeV and background at $\sqrt{s}=10$ TeV and ${\cal L}_{int}=10 ~ ab^{-1}$. Production cross-section for signal (background) is 0.1025 fb (1.44 pb) once the basic selection cuts are imposed.}
    \label{tab:dim6}
\end{table}

\begin{figure}[!hptb]
    \centering
        \includegraphics[width=6.0cm,height=5.5cm]{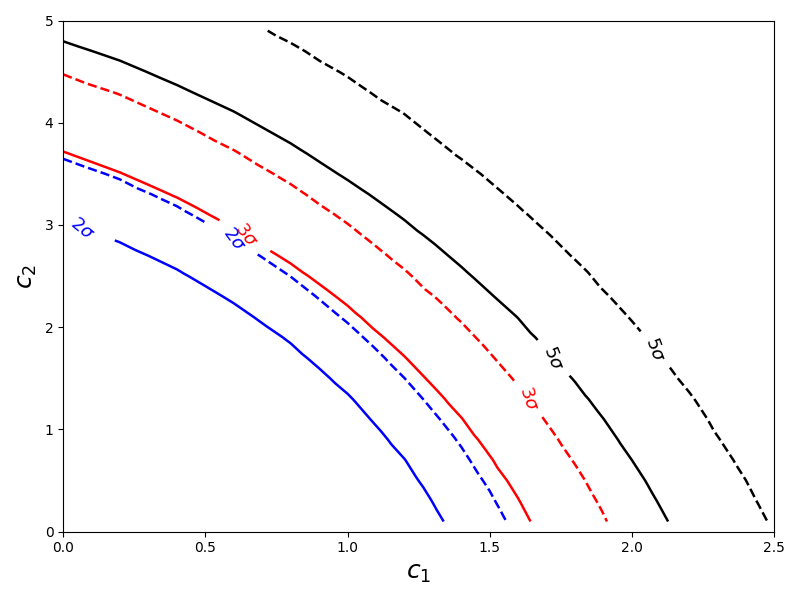}
        \includegraphics[width=6.0cm,height=5.5cm]{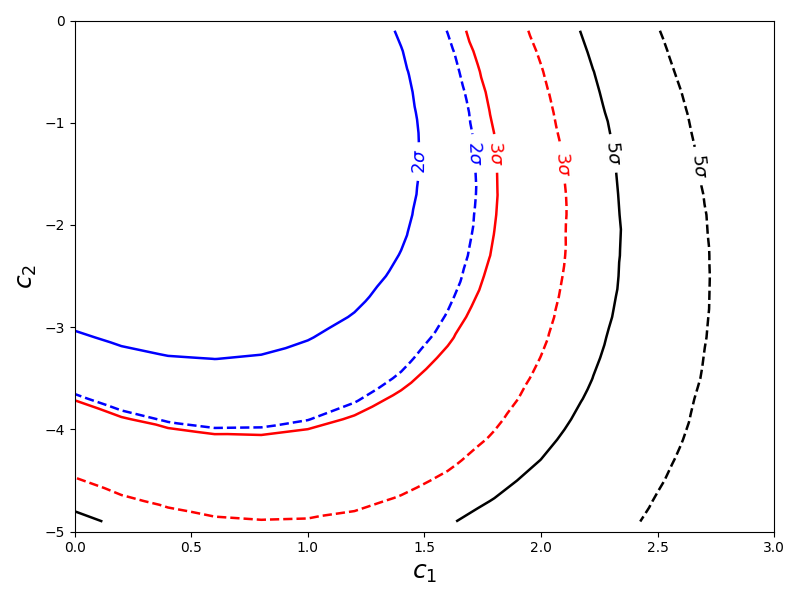}
    \caption{2$\sigma$, 3$\sigma$ and 5$\sigma$ (signal significance) contour plots in the Wilson coefficients $c_1, c_2$ plane, for $m_{\chi} = 1$ TeV (solid), 1.5 TeV (dashed), at $\sqrt{s}=10$ TeV and ${\cal L}_{int}=10 ab^{-1}$.}
    \label{sigma}
\end{figure}

\noindent
We generate events in \texttt{MG5$\_$aMC$\_$v3.5.5}~\cite{Alwall:2014hca,Alwall:2011uj} with the following selection criteria:

$p_T$ of the two muons $>$ 10 GeV and $|\eta|$ of the two muons $<$ 7. 

\noindent
After selecting such events we run the detector simulation via \texttt{Delphes-v3.5.0}~\cite{Selvaggi:2014mya}. We use the Delphes muon collider card, applying 1\% resolution on muon $p_T$. Next we apply the cuts shown in table ~\ref{tab:dim6} sequentially on the aforesaid observables. For $m_\chi = 1$ TeV, $c_1=c_2=2$ and $\Lambda=12$ TeV, we achieve the signal significance of 6.7 $\sigma$. We present the 
cross-sections for three representative masses of the DM particle after applying all the cuts listed in the table \ref{tab1}. Background rate after application of the similar set of cuts are also listed in the same table.

\begin{table}[h]
\centering
\begin{tabular}{|c|c|c|c|}
\hline
\multicolumn{4}{|c|}{Cross-section in fb at $\sqrt{s} = 10 ~\rm TeV$ } \\ \hline
$m_\chi =800$ GeV & $m_\chi =1$ TeV &  $m_\chi =1.2$ TeV & Background \\ \hline
0.058 & 0.0511 & 0.046 & 0.576 \\
\hline
\end{tabular}
\caption{Signal and background cross-sections at $\sqrt{s} = 10 ~\rm TeV$ after application of all the cuts listed in table \ref{tab:dim6}.
$c_1 = c_2 = 2$ and $\Lambda = 12 \rm ~TeV$ has been assumed.}
\label{tab1}
\end{table}

Finally, we perform a scan and present the signal significance ($\sigma)$ contour in the plane of $c_1$ and $c_2$ for two mass points $m_\chi=1$ TeV and 1.5 TeV in Fig.~\ref{sigma}. The left panel of Fig.~\ref{sigma} considers $c_1$ and $c_2$ of same sign and the right panel focuses on the opposite sign for $c_1$ and $c_2$.

\section{Analysis: Inert Doublet Model}


\noindent
Next, we continue our analysis in the context of Inert Doublet Model~\cite{Deshpande:1977rw,Barbieri:2006dq,Cao:2007rm}, where DM is part of a doublet. This model consists of two complex scalar doublets $\Phi_1$ and $\Phi_2$ that can be written as
\begin{align}
    \Phi_1=\begin{pmatrix}
      G^+\\
      \frac{1}{\sqrt{2}}(v+h+iG^0)
    \end{pmatrix}\,,
    \quad \text{and} \quad
    \Phi_2=\begin{pmatrix}
      H^+\\
      \frac{1}{\sqrt{2}}(H+iA)
    \end{pmatrix}\,,
\end{align}

\noindent
Where $\Phi_1$ is SM-like doublet and $\Phi_2$ is inert doublet which does not get a vev and is stabilized by a $Z_2$ symmetry under which $\Phi_2 \rightarrow -\Phi_2$ and all SM fields remain invariant.

We consider $H$ to be the lightest among the three newly added $Z_2$-odd scalars and therefore it serves as a DM candidate here. One can similarly consider $A$ also a stable DM, if $m_A < m_H$. The scalar potential and details of the model can be found in the literature~\cite{Belanger:2015kga,Swiezewska:2012eh,Gustafsson:2012aj,Arhrib:2013ela,Krawczyk:2013jta,Ilnicka:2015jba,Diaz:2015pyv,Belyaev:2016lok,Eiteneuer:2017hoh,Ilnicka:2018def,Kalinowski:2018ylg,Banerjee:2019luv,Tsai:2019eqi,Fabian:2020hny,Kalinowski:2020rmb,Basu:2020qoe,Banerjee:2021xdp,Banerjee:2021hal,Banerjee:2021anv,AbdusSalam:2022idz,Belyaev:2022wrn,Fan:2022dck,Astros:2023gda,Justino:2024etz,Braathen:2024ckk,Abouabid:2025whn,Bal:2025nbu,Ghosh:2025dcv,Gorczyca:2011rs}. 

We summarize the salient features of this model as follows.

\begin{itemize}
\item The particle content of SM is augmented by a DM particle $H$, another neutral scalar $A$, charged scalar $H^{\pm}$.
\item The model parameter space can be spanned by mass of the dark sector scalars $m_H, m_A, m_H^{\pm}$, the DM-SM portal coupling $\lambda_{345}$ and DM self-coupling $\lambda_2$.
\item $\lambda_{345}$ plays a crucial role in DM annihilation and observed relic density as well as direct detection cross-section. 
\end{itemize}

\subsection{Constraints and benchmark selection}

\begin{figure}[!hptb]
    \centering
        \includegraphics[width=8.0cm,height=7.0cm]{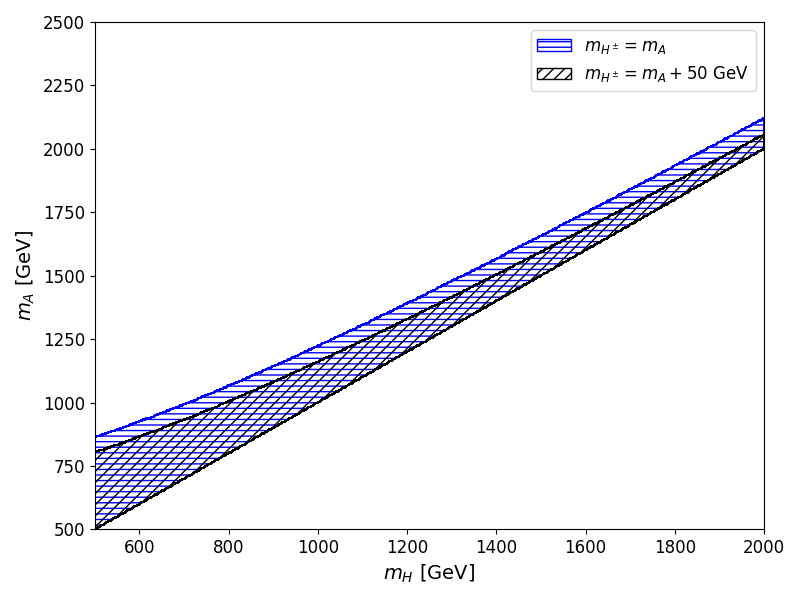}
    \caption{Region allowed by perturbativity, unitarity and boundedness-from-below conditions, in $m_H-m_A$ parameter space. Region with blue horizontal pattern corresponds to $m_H^{\pm}=m_A$, region with black slanted pattern is obtained assuming $m_{H^{\pm}} = m_A + 50$ GeV.}
    \label{unitarity}
\end{figure}

The parameter space of inert doublet model gets restricted by theoretical constraints such as perturbative unitarity and boundedness-from-below. The tree-level perturbative unitarity constrains the mass gap between $m_A$/$m_H^{\pm}$ and the mass of DM $m_H$, so that the quartic couplings of the scalar potential do not become too large~\cite{Gorczyca:2011rs,Dercks:2018wch}. We show in Fig.~\ref{unitarity}, the allowed parameter space in $m_A-m_H$ plane, from all the aforementioned theoretical constraints. The blue horizontal region corresponds to $m_H^{\pm}=m_A$, in order to satisfy the precision electroweak data, i.e. S,T,U parameters~\cite{ParticleDataGroup:2020ssz}. However, a small mass gap between $m_A$ and $m_H^{\pm}$, $\delta m \lesssim 50$ GeV is allowed from the observed S,T,U data~\cite{Grimus:2007if,Grimus:2008nb}. We show the allowed region in black slanted pattern, when $m_H^{\pm}=m_A + 50$ GeV.

Next we discuss the constraints from the DM sector. In this work, we focus on TeV-scale DM, since such heavy particle can be probed at the muon collider. We have checked that for such heavy DM, the relic density contribution decreases significantly and the entire allowed region in Fig.~\ref{unitarity} is relic under-abundant. The dark portal coupling $\lambda_{345}$ is crucial, because it introduces tree-level DM-nucleon scattering. Therefore We should keep $\lambda_{345}$ reasonably low ($\lesssim 10^{-3}$) in order to avoid stringent constraints from direct detection in particular the LZ experiment~\cite{LZ:2024zvo}. 

\begin{figure}[!hptb]
    \centering
        \includegraphics[width=8.0cm,height=7.0cm]{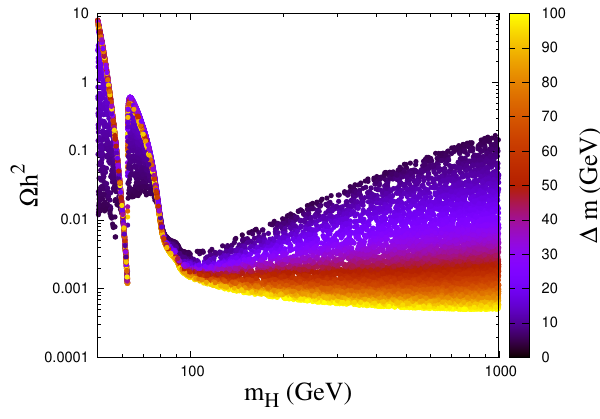}
    \caption{Relic density as a function of DM mass ($m_H$) where $\Delta m = m_A - m_H$ is shown in the color axis. $m_A = m_H^{\pm}$ is assumed. }
    \label{idm_dm}
\end{figure}

One point is worth mentioning here. As is obvious from Fig.~\ref{idm_dm}, the state $H$ is a rather small contributor to the relic density. However, as our estimates below will show, its contribution to the rate for the final state $\mu\mu +$ missing energy can be quite substantial. Therefore, this signal passes off as
one of new physics, through the pair-production  of invisible particles  at a muon collider, however insignificant the participation of such particles in the relic density is as a whole. That is why our illustrative benchmarks and the predictions on their basis are important. 
We have chosen a benchmark that satisfies all the aforementioned constraints and also can be ideal candidate for probing at the muon collider as follows.

$m_H = 1$ TeV, $m_A = 1.2$ TeV, $m_H^{\pm}=$ 1.2 TeV, $\lambda_{345}=0.001, \lambda_2 = 0.1$.  

\noindent
We show the kinematical distributions of various observables for the benchmark and compare with background.

\subsection{Kinematical distributions}

 \begin{figure}[!hptb]
    \centering
        \includegraphics[width=6.0cm,height=4.85cm]{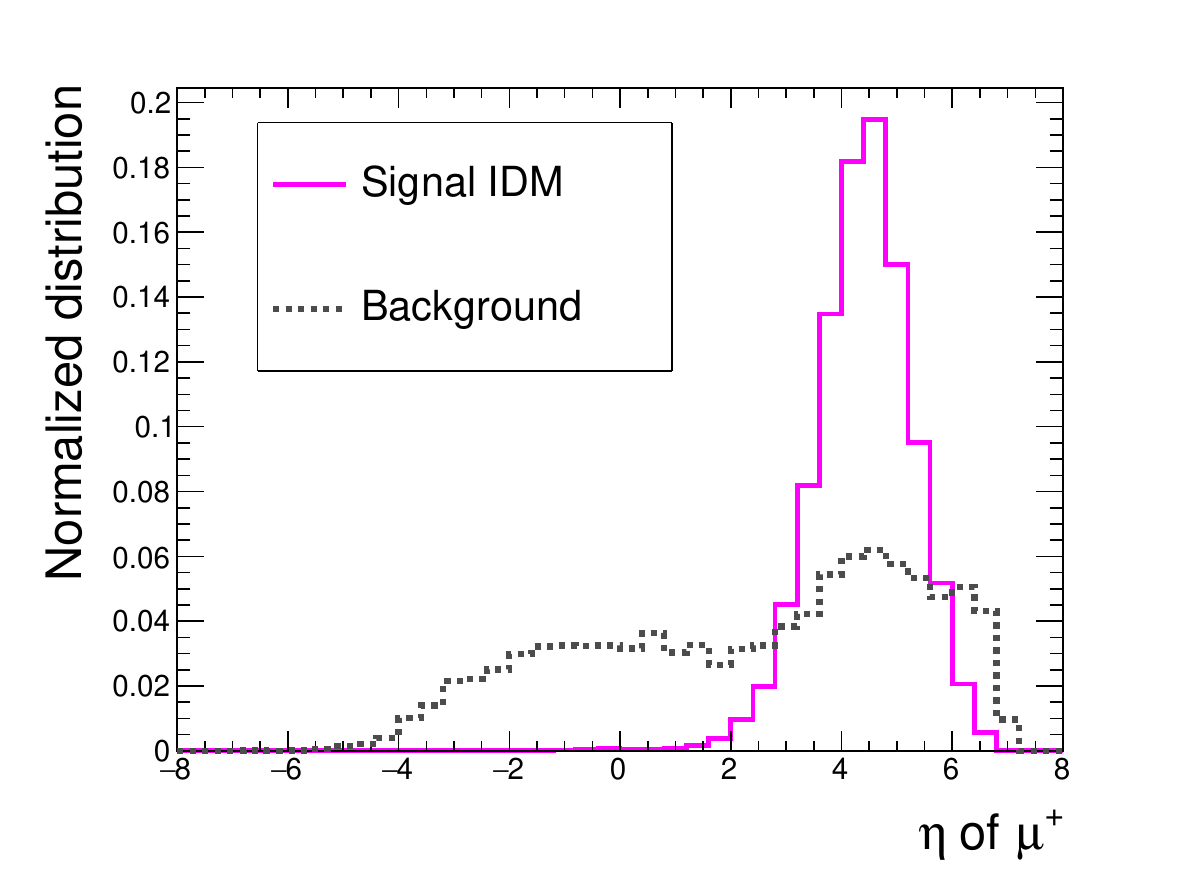}
    \includegraphics[width=6.0cm,height=4.85cm]{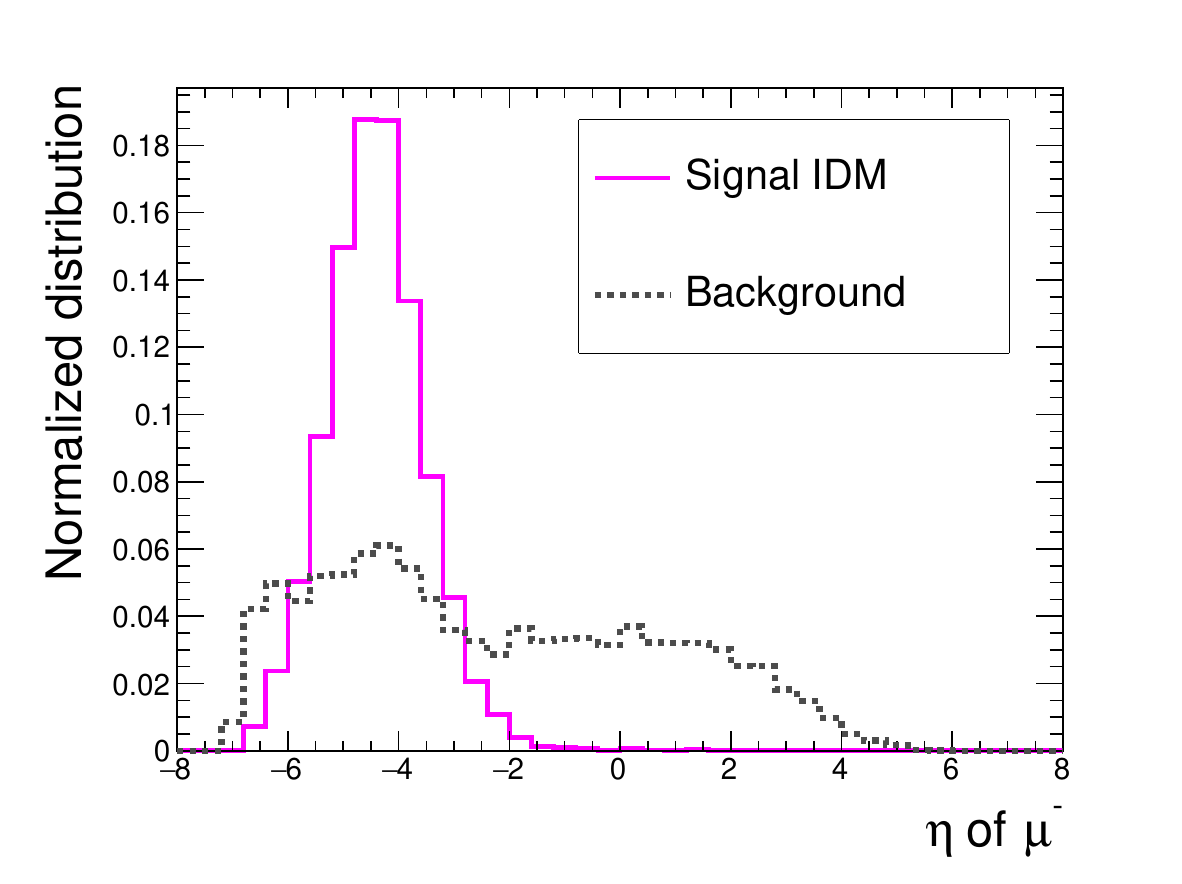}\\
        \includegraphics[width=6.0cm,height=4.85cm]{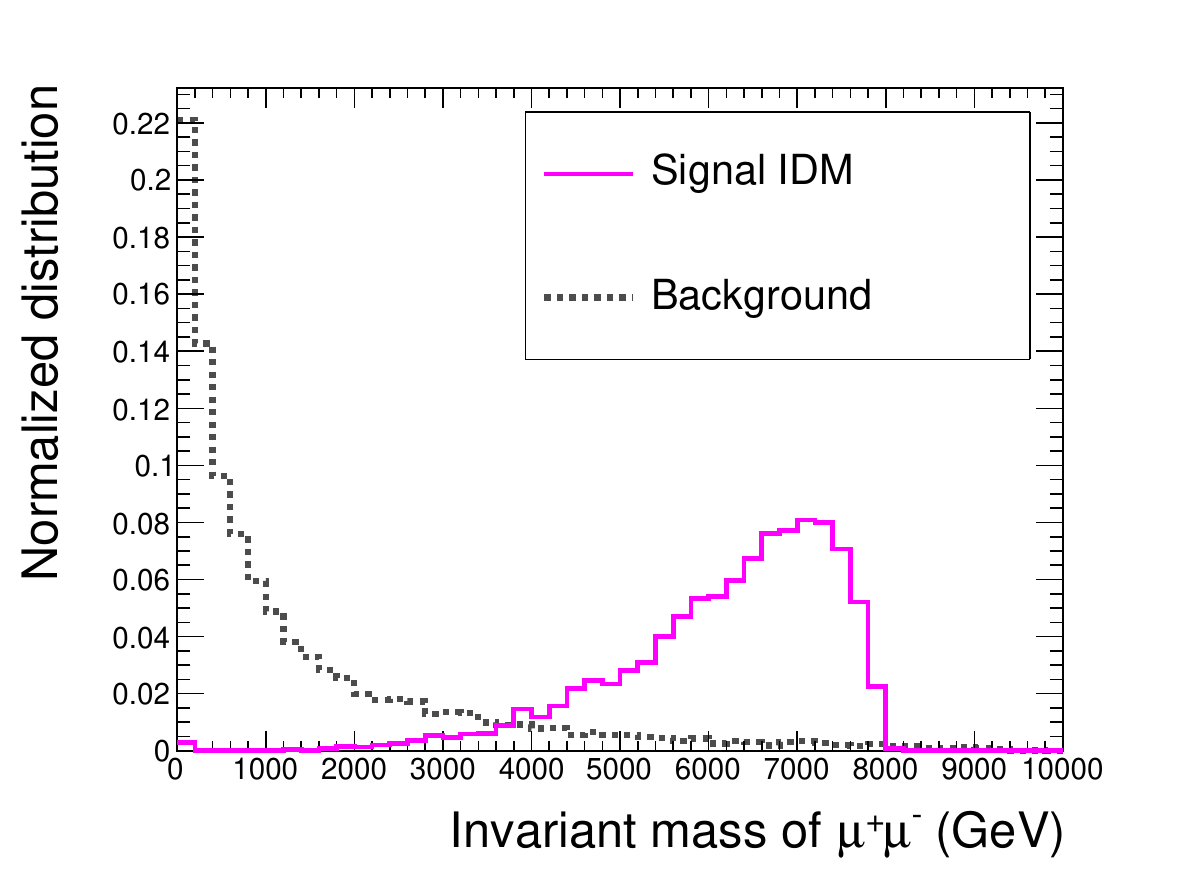}
         \includegraphics[width=6.0cm,height=4.85cm]{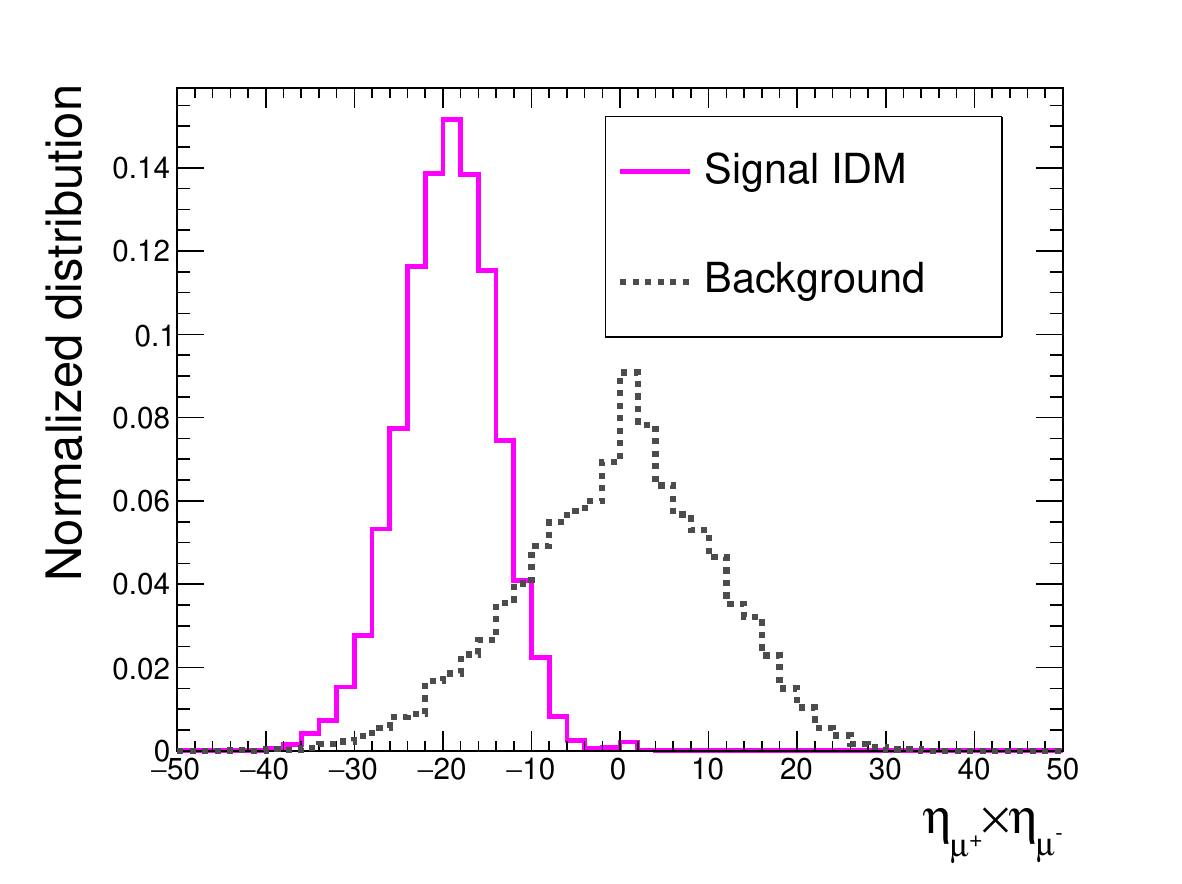}\\
             \includegraphics[width=6.0cm,height=4.85cm]{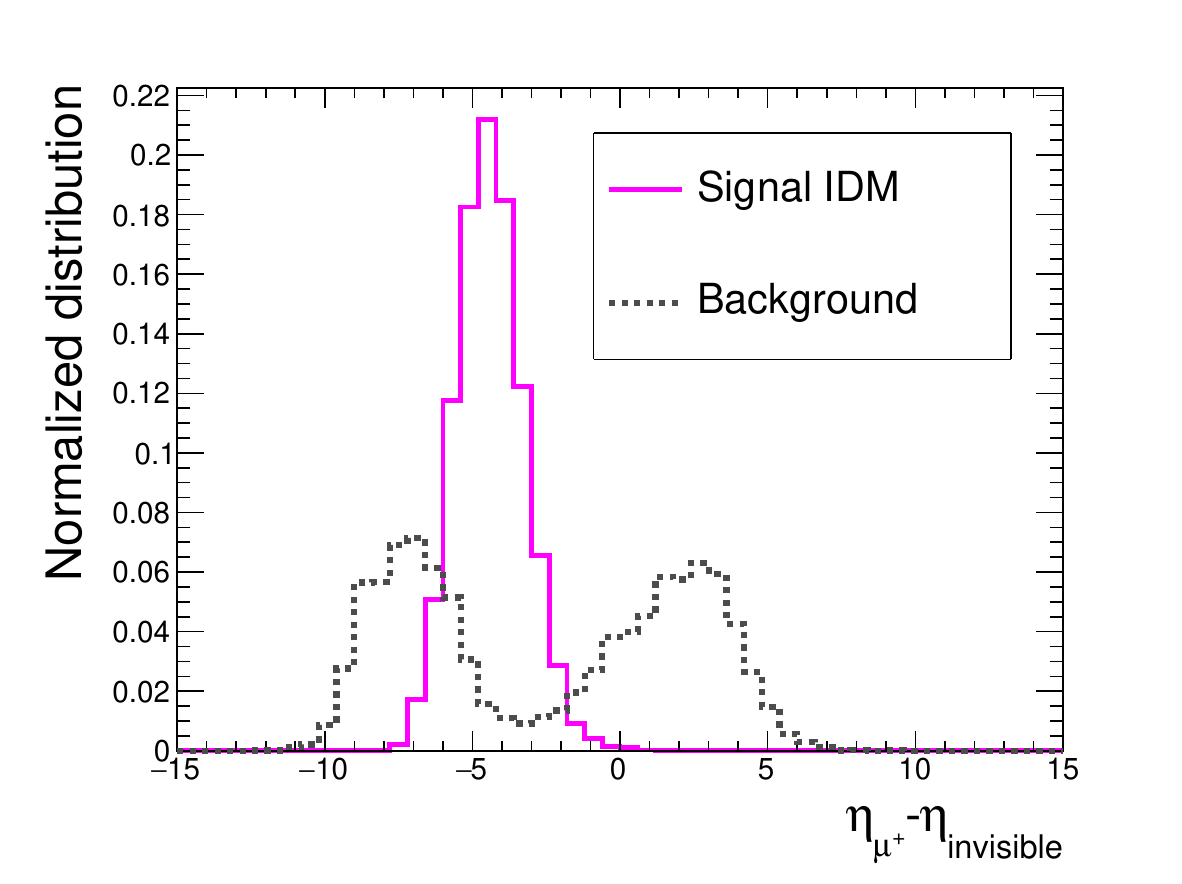}
    \includegraphics[width=6.0cm,height=4.85cm]{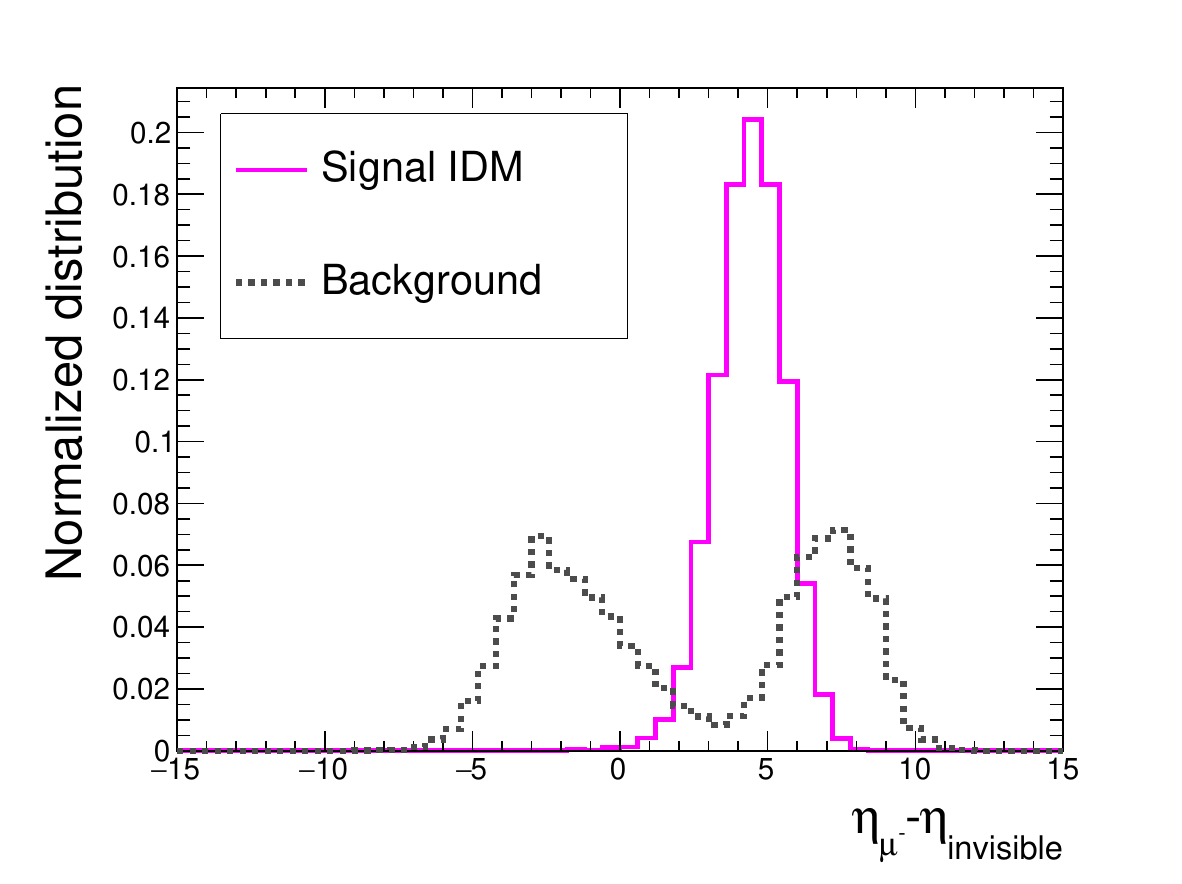}\\
     \includegraphics[width=6.0cm,height=4.85cm]{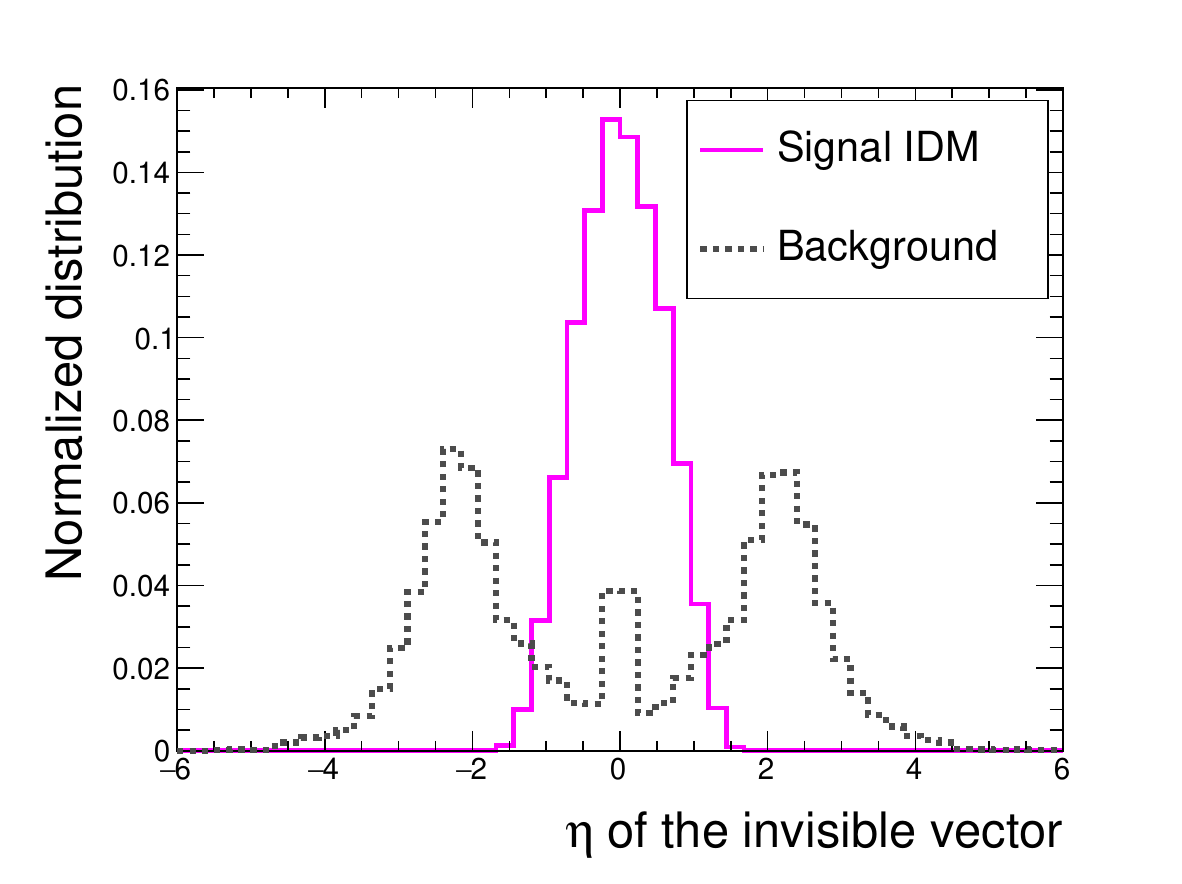}
    \caption{First row: $\eta$ distribution of $\mu^+$ (left) and $\mu^-$ (right), second row: invariant mass of the di-muon pair (left), product of $\eta$ of the di-muon pair (right), third row: difference between $\eta$ of the invisible vector and $\mu^+$ (left), $\mu^-$ (right), fourth row: $\eta$ of the invisible vector. $\sqrt{s}$ = 10 TeV is assumed.}
    \label{dist2}
\end{figure}

In the IDM framework, two forward muons plus missing energy final state arises from the $\mu^+\mu^- \rightarrow  H H \mu^+\mu^-$ process. Pair production of $A$ via $\mu^+\mu^- \rightarrow A A \mu^+\mu^-$, and its subsequent decay $A \rightarrow H Z\;(\rightarrow \nu \bar \nu)$  also contributes to the signal. Although the rate of pair production $A$ and $H$ via VBF are same, the contribution of the former to the signal is suppressed ($\lesssim 1\%)$ due to $Z$-boson decay rate to neutrinos. However for the sake of completeness we have included them in our study.  The major SM background contributing to the same final state comes from $\mu^+\mu^- > \nu\nu \mu^+\mu^-$. The signal process $\mu^+\mu^- \rightarrow  H H \mu^+\mu^-$ is mediated by vector boson
fusion i.e $ZZ$-fusion to a pair of DM. Significant contribution also comes from diagrams involving the dark scalar $A$. Such diagrams involve $ZAH$-coupling. These contributions interfere destructively and result in abysmally small production cross-section when $H$ and $A$ have a small mass-gap. However, we have also seen the mass gap between $H,A$ is constrained by the unitarity bound. In our benchmark we choose the mass gap allowed by the unitarity constraints and also enable substantial signal cross-section. Once more we generate events using \texttt{MG5$\_$aMC$\_$v3.5.5}~\cite{Alwall:2014hca,Alwall:2011uj} with the following selection criteria:

$p_T$ of the two muons $>$ 10 GeV and $|\eta|$ of the two muons $<$ 7. 

\noindent
The final results are obtained on passing each event through \texttt{Delphes-v3.5.0}~\cite{Selvaggi:2014mya}. Here too, we use the Delphes muon collider card, with 1\% resolution on muon $p_T$.

In Fig.~\ref{dist2} (top panel), we first show the pseudo-rapidity distribution of the muons. Since our signal topology is that of vector-boson fusion, the muons are in the forward region (pink curve). The background is more centrally located (black curve). 

We show the invariant mass between the muon pair. The signal shows peak at a higher invariant mass distribution compared to the background (second row left panel). Also, since the two muons are expected to be in the opposite hemisphere owing to the VBF topology, the product of their $\eta$-values are expected to be negative. We show the distribution of $\eta_{\mu^-}\times \eta_{\mu^+}$, which is dominantly negative for signal (second row right panel). One can also see the distribution for signal peaks at a large negative value for the signal. We put a cut on this distribution for effective signal background separation.

We again use the variable $\eta_{invisible}$ introduced earlier in the previous section, where $\eta_{invisible}$ denotes the angle of resultant invisible vector. In case of signal the dark matter pair is expected to be between the forward muon pair. We see this behavior in the third panel of Fig.~\ref{dist2}.

Furthermore, the $\eta_{invisible}$ for signal is concentrated in the central region $|\eta_{invisible}| < 1$, while for background this observable peaks at much larger values. We show this in Fig.~\ref{dist2}. A cut on this variable significantly improve signal selection over background, which we will see below.

\begin{table}[!hptb]
    \centering
    \begin{tabular}{c|c|c}
       Cuts  & No of signal events &  No of Background events \\
        \hline
        Selection cuts  &  630  &  $1.44\times 10^7$ \\
         \hline
         2 isolated muons  & 628 & $1.44\times 10^7$ \\
        \hline
        $2 < |\eta_{\mu}|_{1,2} < 7$ & 562 & $5.2 \times 10^6$\\
        \hline
        $\eta_{\mu^+}\times\eta_{\mu^-} < -10$ & 549 & $2.3 \times 10^6$ \\
        \hline
        Invariant mass of $\mu^+\mu^- > 2$ TeV & 547 & $2 \times 10^6$\\
        \hline
        $\eta_{invisible}-\eta_{\mu^+} < 0$,$\eta_{invisible}-\eta_{\mu^-} > 0$ & 547 & $1.6 \times 10^6$\\
        \hline
        $|\eta_{invisible}| < 1$ & 507 & 8640\\
        \hline
        
   \end{tabular}
   \caption{Cut flow for signal and background at $\sqrt{s} = 10$ TeV and ${\cal L}_{int}=10~ab^{-1}$. Signal BP: $m_H = 1$ TeV, $m_A = m_H^{\pm} = 1.2$ TeV, $\lambda_{345}=0.001, \lambda_2 = 0.1$. Production cross-section for signal (background) is 0.063 fb (1.44 pb) once the basic selection cuts are imposed.}
    \label{idm_cutflow}
\end{table}

\begin{table}[h]
\centering
\begin{tabular}{|c|c|c|c|}
\hline
\multicolumn{4}{|c|}{Cross-section in fb at $\sqrt{s} = 10 ~\rm TeV$ } \\ \hline
$m_H = 800$ GeV,  & $m_H = 1$ TeV,  &  $m_H = 1.2$ TeV,  & Background \\
$\Delta m$ = 260 GeV & $\Delta m$ = 230 GeV & $\Delta m$ = 200 GeV & \\
\hline
0.09 & 0.054 & 0.03 & 0.864 \\ \hline
\end{tabular}
\caption{Signal and background cross-sections in IDM at $\sqrt{s} = 10 ~\rm TeV$ after application of all the cuts listed in table \ref{idm_cutflow}.}
\label{tab2}
\end{table}

\begin{figure}[!hptb]
    \centering
        \includegraphics[width=8.0cm,height=7.0cm]{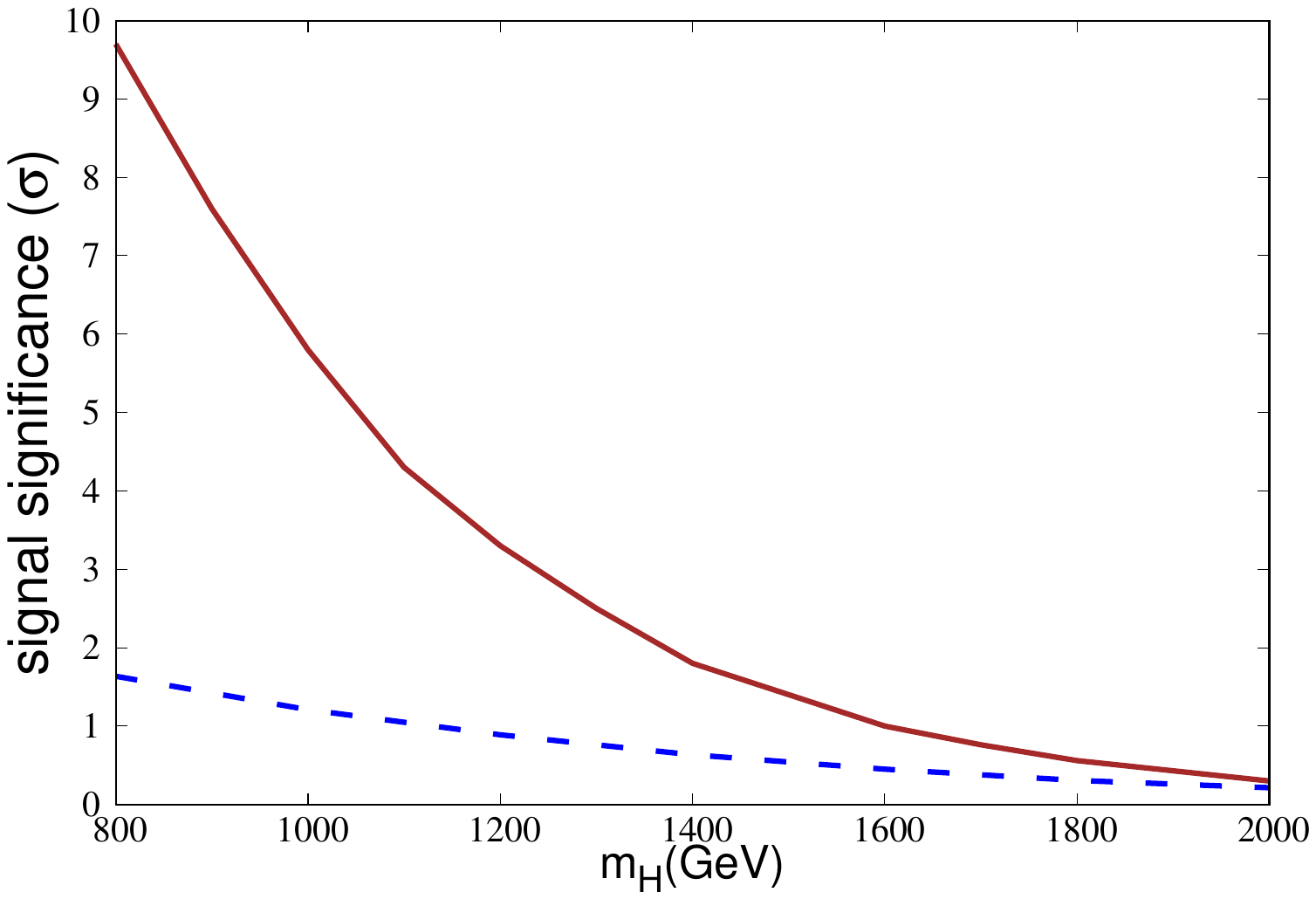}
    \caption{Signal significance as a function of the mass of DM at $\sqrt{s} = 10$ TeV and ${\cal L}_{int}=10~ab^{-1}$. For the maroon plot we have used the maximum possible $\Delta m$ value for each DM mass $m_H$ (allowed range for $\Delta m$ is taken from the region with blue horizontal pattern in Fig.~\ref{unitarity}). In the blue curve, we have chosen $\Delta m$ = 100 GeV for the whole range of $m_H$. }
    \label{scan}
\end{figure}

\noindent
Having discussed the relevant kinematic observables we present the cut-flow for our signal benchmark and SM background in table~\ref{idm_cutflow}. One can calculate the signal significance in this case is 5.5 $\sigma$. As before, in table~\ref{tab2}, we have presented the 
signal cross-sections for three representative values of $m_H$ and $\Delta m$ along with background rate after application of all cuts 
listed in table~\ref{idm_cutflow}.

In Fig.~\ref{scan} we present the signal significance as function of DM mass. We have discussed earlier that the signal cross-section is sensitive to the mass difference between $m_A$ and $m_H$ ($\Delta m$), owing to cancellation between contributing diagrams. For the maroon curve, we have chosen the maximum allowed $\Delta m$ value (from unitarity constraints shown in Fig.~\ref{unitarity}) for each DM mass. In the blue curve, we keep the $\Delta m$ fixed at 100 GeV and show the variation of signal significance as a function of DM mass. The drastic reduction in signal significance in case of $\Delta m =$ 100 GeV, stems from the effective destructive interference between diagrams in case of low mass gap between the DM and its dark sector partners.

\section{Conclusion}

We have investigated the prospect of production and detection of TeV scale dark matter at a muon collider. of c.o.m energy of 10 TeV. In particular we have demonstrated the usefulness of using the Z-fusion channel which leads to a final state $\mu^+\mu^- \chi\chi$, where a $\chi$ is a scalar DM candidate. This channel enables us to utilize the standard criterion of identifying VBF as the principal mechanism. We show that for $\sqrt{s}=10 $ TeV and ${\cal L}_{int} = 10~ab^{-1}$ DM masses of the order of 1 TeV can be identified with sufficient statistical significance. The facility of efficiently detecting energetic forward muons are  crucial for this search. 

As illustration, we have considered (a) a scenario based on dimension-6 interactions and (b) an inert doublet model. We find that with the collider parameters of the above specifications, one can achieve a statistical significance of 6.7 $\sigma$ in case (a) and 5.5 $\sigma$ (b) with reasonable values of the model  parameters.

\section{Acknowledgements}

We thank Alexander Pukhov and Thirthankar Roy Chowdhury for helpful comments.  The work of JL was supported by ANRF grant no-RJF/2025/000544. The work of BM was partially supported by a Raja Ramanna chair position of the Department of Atomic energy, Govt of India.

\bibliographystyle{JHEP}
\bibliography{ref}

\end{document}